\documentclass[preprin,preprintnumbers,eqsecnum,amsmath,amssymb,nofootinbib]{revtex4}
\usepackage[latin9]{inputenc}
\usepackage{amsmath}

\makeatletter

\usepackage{float}
\usepackage{array}
\usepackage{lipsum}
\usepackage{graphicx}
\usepackage{amsmath,amsthm,amsfonts,amssymb}
\usepackage{graphicx}
\usepackage[english]{babel}
\usepackage{color}
\usepackage{tensor}
\usepackage{esint}
\usepackage[dvips]{epsfig}
\usepackage[dvips]{graphicx}
\usepackage{float}
\usepackage{units}
\usepackage{textcomp}
\usepackage{mathrsfs}
\usepackage{amsmath}
\usepackage[makeroom]{cancel}
\usepackage{amssymb}
\usepackage{amsbsy}
\usepackage{amsfonts}
\usepackage{amssymb,mathrsfs,xcolor}
\usepackage{esint}
\usepackage{array}
\usepackage{graphicx}

\usepackage{hyperref}
\hypersetup{
    colorlinks,
    citecolor=blue,
    filecolor=green,
    linkcolor=purple,
    urlcolor=red,
}

\usepackage{slashed}

\newcommand{\ie}{\begin{equation}}
\newcommand{\fe}{\end{equation}}
\newcommand{\se}{\begin{eqnarray}}
\newcommand{\ff}{\end{eqnarray}}

\usepackage{hyperref}
\hypersetup{colorlinks,breaklinks,
			citecolor=[rgb]{0,0.0,1.0},
            urlcolor=[rgb]{0.0,0.0,1.0},
            linkcolor=[rgb]{0,0.5,0.9}}

\begin{document}

\title{Remarks on the effects of the quintessence on regular black holes}

\author{R. V. Maluf$^{1}$}
\email{r.v.maluf@fisica.ufc.br}

\author{C. R Muniz$^{2}$}
\email{celio.muniz@uece.br}
\author{A. C. L. Santos$^{1}$}
\email{alanasantos@fisica.ufc.br}

\affiliation{$^{1}$Universidade Federal do Cear\'{a} (UFC), Departamento de F\'{i}sica, Campus do Pici, Fortaleza - CE, C.P. 6030, 60455-760 - Brazil.\\
 $^{2}$Universidade Estadual do Cear\'{a} (UECE), Faculdade de Educa\c{c}\~{a}o, Ci\^{e}ncias e Letras de Iguatu, Av. D\'{a}rio Rabelo, s/n, Iguatu-CE, 63.500-000 - Brazil.}

\date{\today}

\begin{abstract}
We present the generalization of a regular black hole surrounded by quintessence, considering Bardeen-like solutions added with this fluid. We also compare our solution particularized to the Bardeen one with those found in the recent literature, showing that some of them are inconsistent since they add quintessence ``by hand'', neglecting also the interesting behavior of the complete solution nearby the origin. Then, we present the correct way to implement the quintessential fluid in more general four-dimensional regular black hole geometries and explore some of its consequences, such as removing the black hole regularity at the origin.
\end{abstract}


\maketitle

\section{Introduction}

Quintessence is a kind of exotic matter that can play a fundamental role in the late-time accelerated expansion of the Universe \cite{Shinji} (and references therein). The current observations point to that the state parameter corresponding to this form of dark energy, if it there exists, is restricted to the interval $-1<\omega<-1/3$, with $\omega=-1$ associated with the cosmological constant. Unlike the latter, quintessence can be variable in space and time, playing therefore a role in Astrophysics. Thus, in the black hole geometries, for instance, the effect of quintessence was analyzed in several scenarios \cite{Kiselev,Ghosh,Muniz1,Jeff1,Saleh}. Notwithstanding, according to \cite{Visser2}, it is more appropriate to consider these black holes as being surrounded by an average cosmological anisotropic fluid, which is technically different from quintessence \cite{Lobo1}.

In particular, the influence of that fluid on the geometry of Bardeen's regular black hole \cite{Bardeen,Eloy,Maluf} was examined firstly in \cite{Ghaderi1,Ghaderi2}. However, these works exhibit some inconsistencies that need to be clarified and corrected, and the present work seeks to do this. Firstly, the inclusion of the quintessence term in the geometry of the Bardeen regular black hole was made ``by hand'' through the asymptotic expansion of its metric, according to Eqs. (9) and (10) presented in \cite{Ghaderi1}. In fact, such an expansion yields a metric that can be any other, as that associated with the quantum corrected Schwarzschild black hole \cite{Bravo}, for instance, which is completely different from the complete Bardeen solution. Thus, the very interesting characteristic of the object nearby or at the origin becomes excluded from the investigation.

The correct form of introducing quintessence in the Bardeen regular black hole geometry was recently made in the literature \cite{Rodrigues}. Thus, to avoid overlap and redundancy, we will make the analysis by considering the quintessential fluid in the spacetime of more general regular black holes, which still remains original. In the sequel, we particularize it to the Bardeen solution. Henceforth, we will assume $c=G=1$. \\

\section{Generalized regular black hole surrounded by quintessence}

The generalized Bardeen-like regular black hole solution is given by \cite{Lobo}

\begin{equation}\label{quint}
ds^2=-\left[1- \frac{2m(r)}{r}\right]dt^2+\left[1- \frac{2m(r)}{r}\right]^{-1}dr^2+r^2(d\theta^2+\sin^2{\theta}d\phi^2),
\end{equation}
where, in the metric function $f(r)=1-2m(r)/r$, we have
\begin{equation}
m(r)=\frac{M r^{k+1}}{(r^{2n}+g^{2n})^{(k+1)/(2n)}},
\end{equation}
with $g$ being the parameter associated to the source of the Bardeen-like (BL) black hole, $M$ the ADM mass, and $k,n$ positive integers. The effective profile of the energy density that generates this BL geometry, which could in turn be derived from the gravity coupled to a nonlinear electrodynamics \cite{Rodrigues}, is found from 00-component of Einstein's equations, given by
\begin{equation}\label{zero-zero}
\frac{r f'(r)+f(r)-1}{r^2}=-\kappa \rho
\end{equation}
where $\kappa=8\pi$, with
\begin{equation}
\rho_{BL}=\frac{2 (k+1) M g^{2 n} r^{k-2}}{8\pi\left(g^{2 n}+r^{2 n}\right)^{\frac{k+2 n+1}{2 n}}}.
\end{equation}
On the other hand, the density profile associated to Kiselev's (K) black hole, is given by \cite{Kiselev}
\begin{equation}
\rho_K=\frac{3c\omega_q}{2r^{3(\omega_q+1)}},
\end{equation}
where $-1<\omega_q<-1/3$, and, therefore, $c<0$ such that the energy density is positive. Thus, in order to have Bardeen-like black hole surrounded by quintessence, we should combine these densities in the simplest form $\rho=\rho_{BL}+\rho_{K}$ and plug it into Eq. (\ref{zero-zero}). Fortunately, the differential equation (\ref{zero-zero}) is linear in $f(r)$, then we can simply add BL and K black hole solutions, obtaining thus
\begin{equation}\label{solutionBLK}
f(r)=1- \frac{2M r^{k}}{(r^{2n}+g^{2n})^{(k+1)/(2n)}}-\frac{a} {r^{3\omega_q+1}},
\end{equation}
where we have adopted $-c=a\geq0$. Notice that this solution  asymptotically corresponds to the Schwarzschild black hole surrounded by quintessence \cite{Kiselev}. Moreover, by particularizing Eq. (\ref{solutionBLK}) to $n=1$ and $k=2$, we obtain the correct Bardeen-Kiselev (BK) solution recently studied in \cite{Rodrigues}, without the cosmological constant term.

In Fig. 1 we depict the metric function $f(r)$ for this solution with $\omega_q=-2/3$, comparing it with Schwarzschild-Kiselev (SK) and Reissner-Nordstr\"{o}m-Kiselev (RNK) black hole solutions, in order to analyse their horizons.
\begin{figure}[!ht]
    \centering
    \begin{minipage}{0.45\linewidth}
        \centering
        \includegraphics[width=0.9\textwidth]{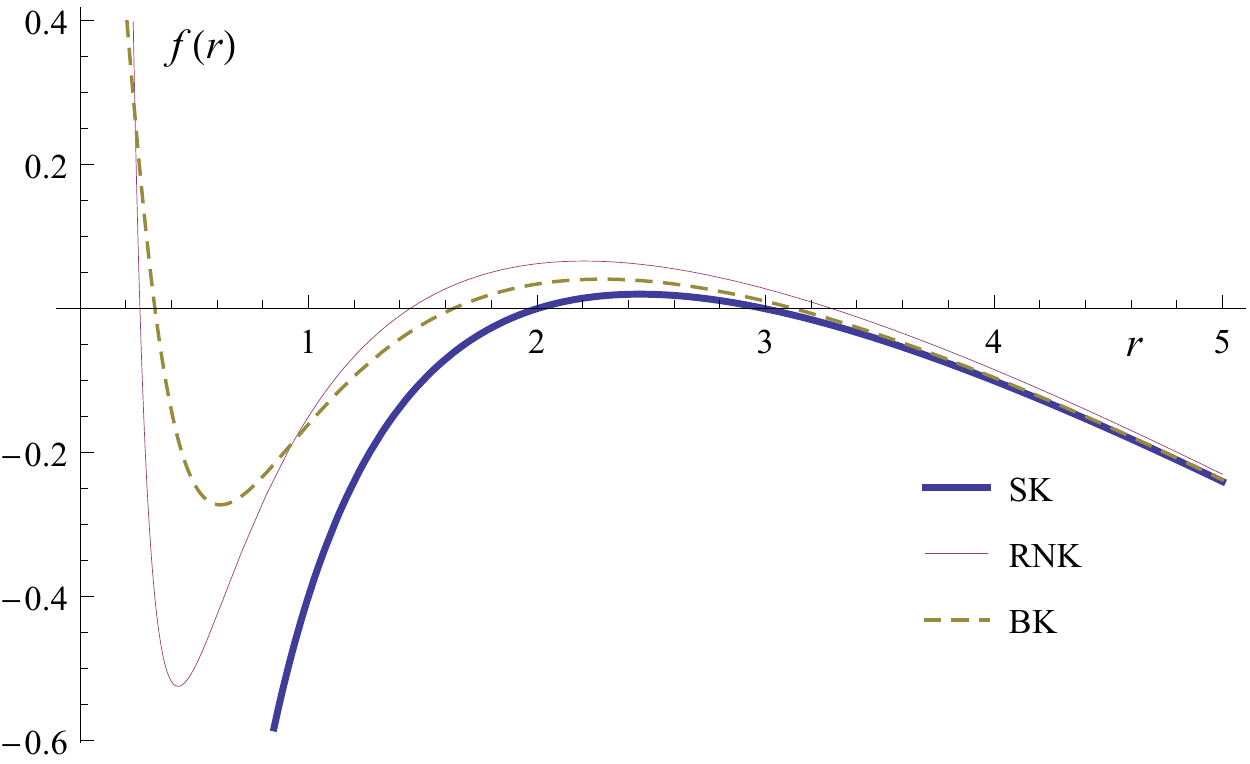}
            \end{minipage}\hfill
    \begin{minipage}{0.45\linewidth}
        \centering
        \includegraphics[width=0.9\textwidth]{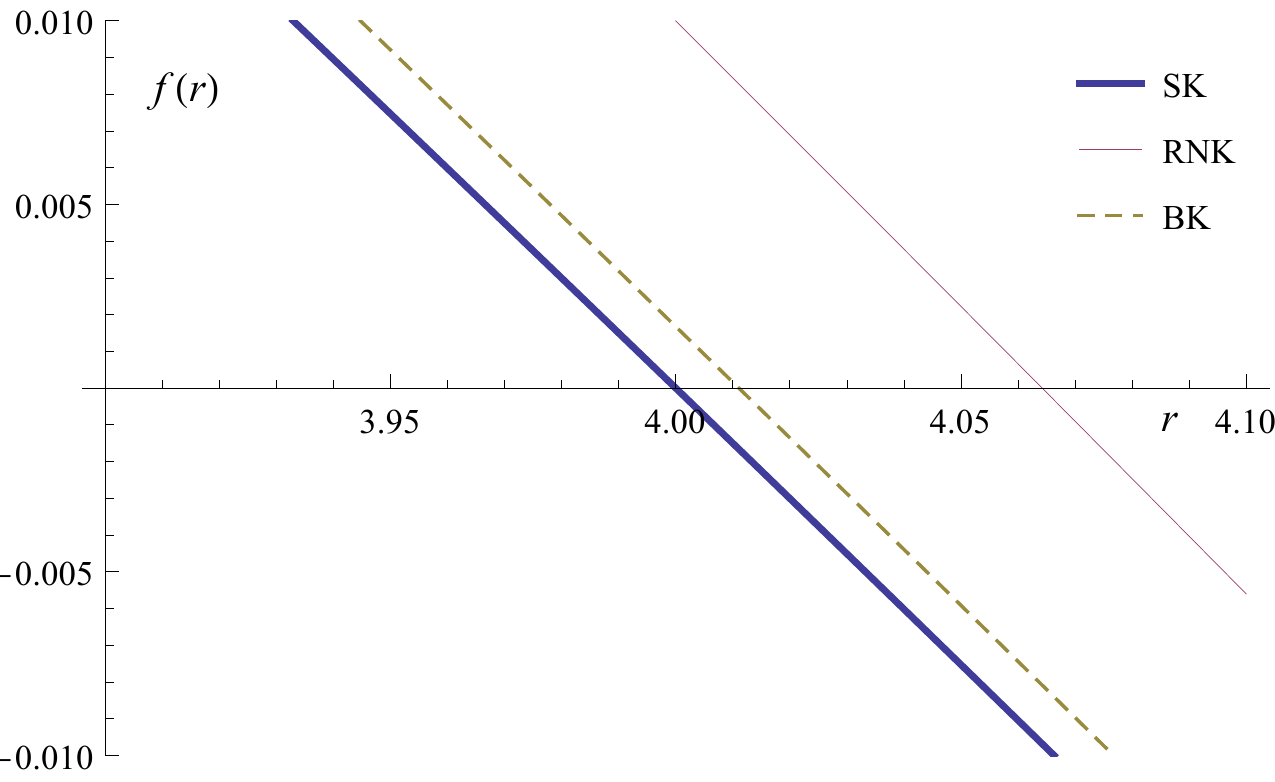}
                \end{minipage}
        \caption{The metric function in terms of the radial coordinate $r$ for the three black hole solutions. The parameter settings are $\omega_q=-2/3$, $M=0.4$, $q=0.4$ (charge of RNK black hole), $g=0.3$ and $a=0.2$, in Planck units.}
    \label{figurasminipg1}
    \end{figure}
Differently from the result presented in \cite{Ghaderi1,Ghaderi2}, the BK black hole solution comprises up to four horizons, from which three we can see in the left panel of Fig. 1, due to the chosen parameters. These horizons correspond to four real roots of the equation $f(r)=0$, an involved expression that does not allow that one computes analytically these horizons, as it was made in the mentioned papers. On the other hand, the RNK solution can have up to three horizons, and SK black hole can only present up to two horizons. It is worth notice that the addition of quintessence to the general solution of a regular black hole makes the object no longer regular at the origin, at least for $\omega\geq -1$ or $k< 2$, although $f(r)$ be defined at that point. In fact, the Ricci curvature scalar nearby the origin, for $\omega=-2/3$, $n=1$, and $k=2$ results in
   \begin{equation}
   R\approx \frac{6 a}{r}+\frac{24 M}{g^3}.
   \end{equation}
If we remove the quintessence, by making $a\to 0$, the regularity is retrieved.

Furthermore, the right panel of Fig. \ref{figurasminipg1} shows the behaviour of the outer horizon of the analysed black holes. A simple inspection of the metric function $f(r)$ for the general Bardeen-Kiselev solution ({\it i.e.}, for any $k,n$) is always greater than the one of the SK black hole, provided $g^{2n}>0$, revealing thus that the outer (inner) horizon of the former is always larger (smaller) than the horizon associated with the latter, a result that is completely different from that is shown in \cite{Ghaderi1,Ghaderi2}. This is more a problem that occurs due to one analyzing only the asymptotic expansion of the solution.

From the foregoing, the authors of reference \cite{Ghaderi1,Ghaderi2} spoiled their work from the beginning, and we fear it is irremediable, since the subsequent thermodynamic analysis was made based on the incomplete solution for the Bardeen-Kiselev regular black hole that they have obtained. However, such physical implications of the studied model might be considered if the original results are fitted to the complete Bardeen's solution, leaving as future perspective its generalization to the Bardeen-like solution with quintessence presented here.

\begin{acknowledgments}
The authors thank the Funda\c{c}\~{a}o Cearense de Apoio ao Desenvolvimento
Cient\'{i}fico e Tecnol\'{o}gico (FUNCAP), the Coordena\c{c}\~{a}o de Aperfei\c{c}oamento de Pessoal de N\'{i}vel Superior (CAPES), and the Conselho Nacional de Desenvolvimento Cient\'{i}fico e Tecnol\'{o}gico (CNPq), Grants no 311732/2021-6 (RVM) and no 308168/2021-6 (CRM) for financial support.

\end{acknowledgments}


\end{document}